\documentclass{svproc}
\usepackage{url}

\usepackage{graphicx}

\newcommand{\be}{\begin{equation}}
\newcommand{\ee}{\end{equation}}

\newcommand{\bea}{\begin{eqnarray}}
\newcommand{\eea}{\end{eqnarray}}

\usepackage{here}

\begin{document}

\mainmatter              
\title{The Impact of Trump-Era Tariffs on Financial Market Efficiency}
\titlerunning{The Impact of Trump-Era Tariffs}  
%
\author{Tetsuya Takaishi}
\authorrunning{Tetsuya Takaishi} 
%
%
\institute{Hiroshima University of Economics, Hiroshima 731-0192, Japan,\\
\email{tt-taka@hue.ac.jp}
}

\maketitle              

\begin{abstract}
This study examines the effects of Trump-era tariffs on financial market efficiency by applying multifractal detrended fluctuation analysis to the return and absolute return time series of six major financial assets: the S\&P 500, SSEC, VIX, BTC/USD, EUR/USD, and Gold. Using the Hurst exponent $h(2)$ and multifractal strength, we assess how market dynamics responded to two major global shocks: the COVID-19 pandemic and the implementation of the Trump tariff policy in 2025. The results show that COVID-19 induced substantial changes in both the Hurst exponent and multifractal strength, particularly for the S\&P 500, BTC/USD, EUR/USD, and Gold. In contrast, the effects of the Trump tariffs were more moderate but still observable across all examined time series. The Chinese market index (SSEC) remained largely unaffected by either event, apart from a distinct response to domestic stimulus measures. In addition, the VIX exhibited anti-persistent behavior with $h(2) < 0.5$, consistent with the rough volatility framework. These findings underscore the usefulness of multifractal analysis in capturing structural shifts in market efficiency under geopolitical and systemic shocks.
\end{abstract}

\section{Introduction}
The Efficient Market Hypothesis (EMH), introduced by Eugene Fama in 1970, is a cornerstone of modern financial economics~\cite{Fama1970efficient}. It posits that financial markets are \emph{informationally efficient}, meaning that asset prices fully reflect all available information at any given time. Consequently, it is impossible to consistently earn returns that outperform the market through information-based trading strategies. Under the assumption that price movements follow a random process, the EMH implies that financial time series should exhibit randomness. Therefore, evaluating the degree of randomness in market data serves as a practical approach to assessing market efficiency.

The randomness of a time series can be quantified by the Hurst exponent~\cite{hurst1951long}. When $H = 0.5$, the series is random; values of $H > 0.5$ indicate persistence, whereas $H < 0.5$ indicate anti-persistence. Previous studies have shown that in developed stock markets, the Hurst exponent 
of return time series typically approaches 0.5, suggesting market efficiency, while in emerging markets it often exceeds 0.5, implying inefficiency. Moreover, certain developed markets exhibit $H < 0.5$, reflecting anti-persistent dynamics~\cite{MATTEO2005827,takaishi2022time}.

Market efficiency can fluctuate due to various factors. For example, in the early stages of the Bitcoin market, the Hurst exponent was found to be below 0.5, exhibiting anti-persistent behavior. As the market matured, however, the exponent approached 0.5, suggesting a gradual movement toward efficiency~\cite{takaishi2019market}. Similarly, the power-law exponent $\alpha$ characterizing the tails of Bitcoin return distributions was initially around 2, smaller than the canonical value of $\alpha = 3$ known for traditional assets~\cite{gopikrishnan1998inverse,gopikrishnan1999scaling}, but has since converged toward $\alpha \approx 3$ as the market developed~\cite{easwaran2015bitcoin,drozdz2018bitcoin,takaishi2021recent}.
The tail behavior of return distributions affects the values of risk measures such as Value at Risk (VaR) and Conditional Value at Risk (CVaR). When the return distribution follows a power-law, it is known that the ratio between VaR and CVaR is determined solely by the power-law exponent\cite{takaishi2023properties}. This implies that, under power-law distributions, VaR and CVaR are no longer independent risk metrics.

In equity markets, the existence of the leverage effect—where volatility tends to increase more following negative returns than after positive ones—is well established\cite{Black1976,Christie1982stochastic,wu2001determinants}. In contrast, early studies on the Bitcoin market have reported a reverse leverage effect, in which volatility responds more strongly to positive returns\cite{bouri2016return,stavroyiannis2017dynamic}. However, this effect has shown temporal variation within the Bitcoin market, and recent research suggests that its intensity has diminished in recent years\cite{takaishi2018,takaishi2021time}.

Market efficiency is also sensitive to major global events. For instance, the COVID-19 pandemic—officially declared by the World Health Organization (WHO) on 11 March 2020—had a profound impact on global financial systems. Numerous studies have documented variations in both market efficiency and multifractal strength in response to the pandemic (see, e.g.,~\cite{aslam2020efficiency,fernandes2022evaluating,wang2020analysis,shen2022multifractal,mensi2021does,alijani2021fractal,zitis2023investigating,raza2024multifractal,erer2023aggregate,aslam2020evidence,mensi2020impact,ameer2023impact,saadaoui2023skewed,arouxet2022covid,takaishi2025impact,gao2024impact}). However, the magnitude and nature of these impacts differ across markets and data types\cite{takaishi2025impact}. In China, for example, the effect of COVID-19 on the efficiency of return time series appears limited, while volatility-based series exhibit distinct patterns of response.

The second Trump administration officially commenced on January 20, 2025. Shortly thereafter, on April 2, 2025, a new trade policy—widely referred to as the \emph{Trump Tariffs}—was introduced. Under this policy, a baseline tariff of 10\% was imposed on imports from nearly all countries beginning April 5, 2025. The announcement triggered significant volatility in global financial markets, with the S\&P 500 index falling by approximately 11\% within two days. Following subsequent bilateral negotiations, tariff rates were adjusted on a country-by-country basis, and the policy’s effects have continued to persist.

This study investigates the impact of the Trump Tariffs on financial market efficiency. To quantify efficiency, we employ the multifractal detrended fluctuation analysis (MFDFA) framework~\cite{kantelhardt2002multifractal} to estimate the generalized Hurst exponent. The analysis covers periods affected by both COVID-19 and the Trump Tariffs, allowing for a comparative assessment of their respective impacts. Furthermore, by analyzing both return and volatility (absolute return) time series, we evaluate how the type of data influences the degree of observed effects.

\section{Methodology}

To examine the multifractal properties of financial time series, we employ the MFDFA~\cite{kantelhardt2002multifractal}. The procedure is described as follows.

(i) We first construct the profile $Y(i)$ as
\be
Y(i) = \sum_{j=1}^i \left[r(j) - <r> 
\right],
\ee
where $<r>$ denotes the mean of the return series.

(ii) The profile $Y(i)$ is then divided into $N_s$ non-overlapping segments of equal length $s$, where $N_s \equiv \mathrm{int}(N/s)$. Because the total length of the series is not always a multiple of $s$, a short residual segment may remain at the end. To make full use of the data, the same segmentation procedure is repeated starting from the end of the profile. Thus, a total of $2N_s$ segments are obtained.

(iii) For each segment $\nu$ ($\nu = 1, \dots, N_s$), we compute the local variance as
\be
F^2(\nu, s) = \frac{1}{s} \sum_{i=1}^s \left[Y((\nu-1)s+i) - P_\nu(i)\right]^2,
\ee
and similarly, for $\nu = N_s+1, \dots, 2N_s$,
\be
F^2(\nu, s) = \frac{1}{s} \sum_{i=1}^s \left[Y(N-(\nu-N_s)s+i) - P_\nu(i)\right]^2,
\ee
where $P_\nu(i)$ denotes the fitting polynomial used to remove the local trend in segment $\nu$. The polynomial is defined as
\be
P_\nu(i) = \sum_{k=0}^p a_k i^k,
\ee
and in this study, we employ a cubic-order polynomial ($p=3$).

(iv) The $q$th-order fluctuation function is then obtained by averaging over all segments:
\be
F_q(s) = \left\{ \frac{1}{2N_s} \sum_{\nu=1}^{2N_s} [F^2(\nu,s)]^{q/2} \right\}^{1/q}.
\label{eq:FL}
\ee

(v) The generalized Hurst exponent, denoted as $h(q)$, is estimated from the scaling behavior of $F_q(s)$. For a time series exhibiting long-range power-law correlations, $F_q(s)$ scales as
\be
F_q(s) \sim s^{h(q)},
\label{eq:asympto}
\ee
for sufficiently large $s$.

The singularity spectrum $f(\alpha)$, which characterizes the multifractal nature of the time series, can be derived from $h(q)$ using the Legendre transform:
\be
f(\alpha) = q[\alpha - h(q)] + 1,
\ee
where $\alpha$ represents the H\"{o}lder exponent or singularity strength, defined as
\be
\alpha = h(q) + qh'(q).
\ee

The Hurst exponent, a measure of market efficiency, is obtained from $h(2)$. In this study, the range of $q$ is restricted to $q \in [-5, 5]$, because for large $|q|$ the moments in the fluctuation function may diverge, leading to numerical instability in the estimation of $h(q)$~\cite{Jiang-Xie-Zhou-Sornette-2019-RPP}.

Finally, the degree of multifractality is quantified by the width of the generalized Hurst exponent as
\be
\Delta h(q) = h(-q) - h(q),
\label{eq1}
\ee
where $q \neq 0$. A Gaussian time series exhibits monofractal behavior with $\Delta h(q) = 0$, while a nonzero $\Delta h(q)$ indicates multifractality. Therefore, the magnitude of $\Delta h(q)$ can be interpreted as a proxy for the degree of market inefficiency.

\section{Data}
\begin{table}[h]
\centering
\caption{Correspondence between abbreviations and market indicators.}
\begin{tabular}{|c|c|}
\hline
Abbreviation & Market indicator \\
\hline
S\&P500 & Standard \& Poor’s 500 Composite Stock Price Index \\
SSEC & Shanghai Stock Exchange Composite Index \\
VIX & CBOE Volatility Index \\
BTC/USD & Exchange rate between Bitcoin and the US Dollar \\
EUR/USD & Exchange rate between the Euro and the US Dollar \\
Gold & Gold Futures \\
\hline
\end{tabular}
\label{tab:label_value}
\end{table}
We obtain the daily closing prices $P(t)$ for six major financial assets: S\&P 500, SSEC, VIX, BTC/USD, EUR/USD, and Gold. Table~\ref{tab:label_value} summarizes the abbreviations used for each market indicator. The data are sourced from \url{https://www.investing.com} (accessed on 27 September 2025) and cover the period from 1 January 2014 to 26 September 2025, encompassing both the COVID-19 pandemic and the implementation of the Trump tariff policy.

\begin{figure}[H]
\centering
\includegraphics[width=9.5cm]{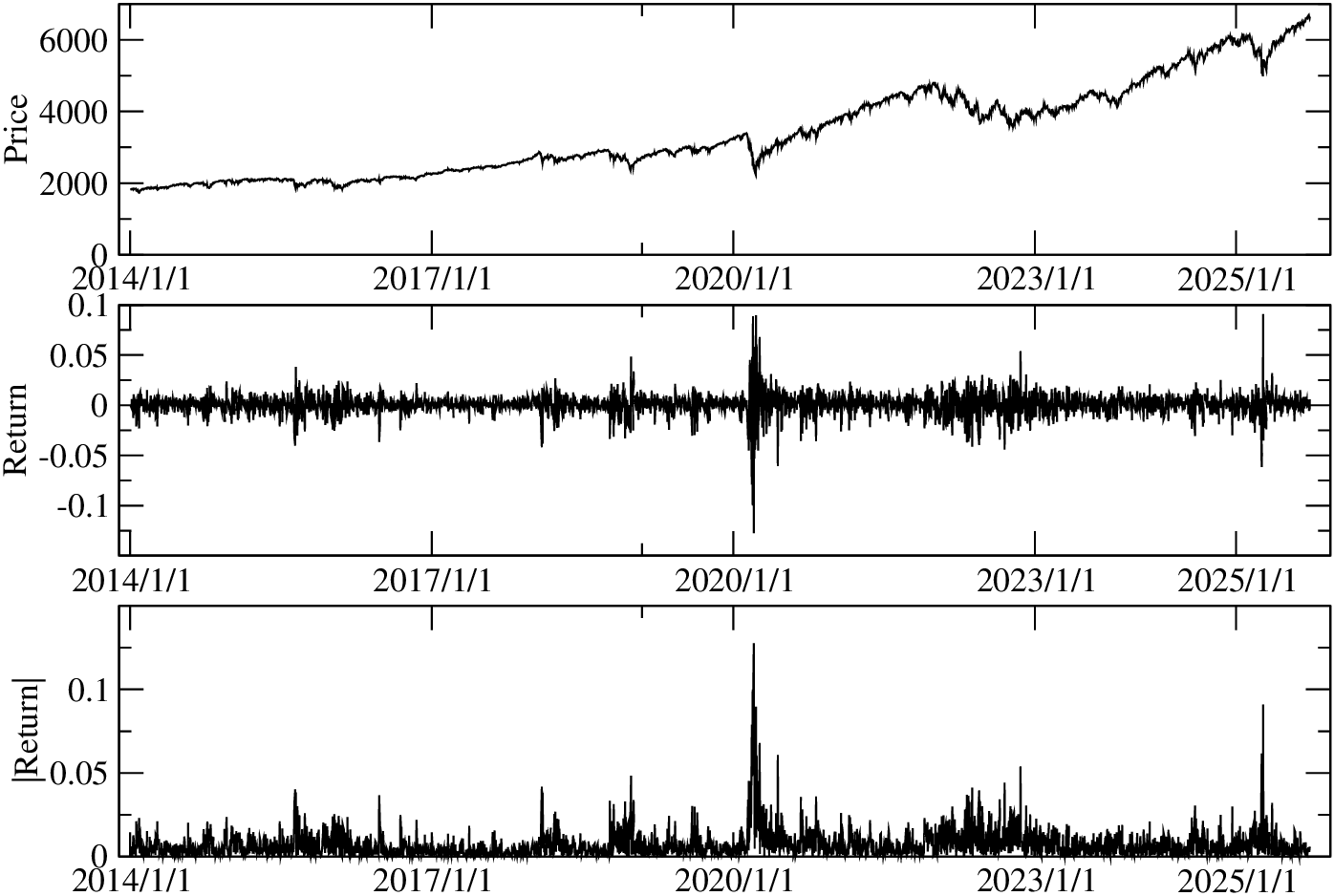}
\caption{Time series of price, return, and absolute return for the S\&P 500.}
\label{fig1}
\end{figure}

\begin{figure}[H]
\centering
\includegraphics[width=9.5cm]{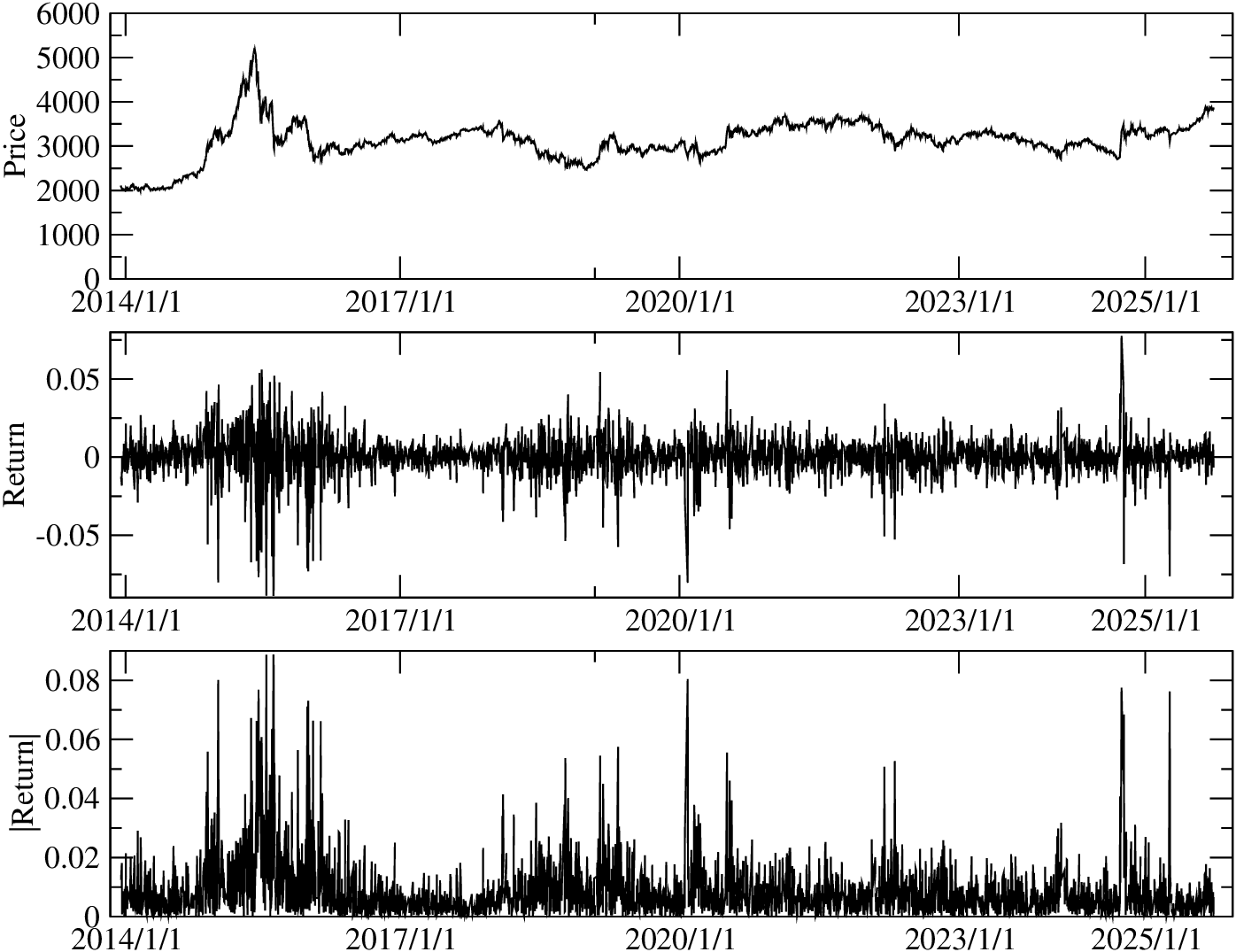}
\caption{Time series of price, return, and absolute return for the SSEC.}
\label{fig:SSEC}
\end{figure}

\begin{figure}[H]
\centering
\includegraphics[width=9.5cm]{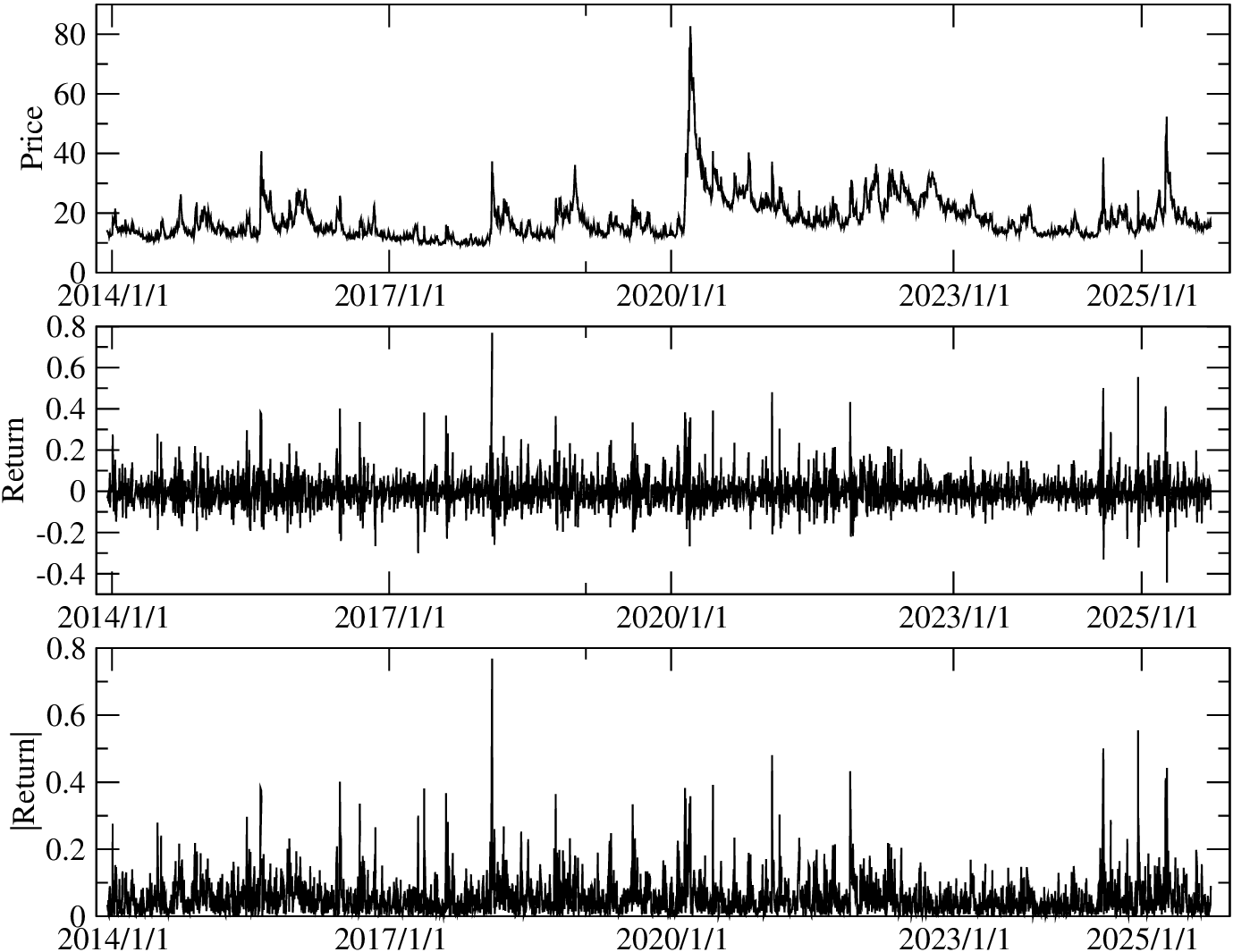}
\caption{Time series of price, return, and absolute return for the VIX.}
\label{fig:VIX}
\end{figure}
The logarithmic return $r(t)$ is defined as
\be
r(t) = \log P(t) - \log P(t-1),
\ee
where $P(t)$ represents the asset price at time $t$. The absolute return $|r(t)|$ is used as a proxy for volatility, which is known to exhibit long-memory characteristics~\cite{ding1993long}.

As representative examples, Figures~1–3 show the time series of prices, returns, and absolute returns for the S\&P 500, SSEC, and VIX, respectively.

\section{Empirical Results}

The generalized Hurst exponent $h(q)$ is estimated using the MFDFA\cite{kantelhardt2002multifractal}, applied to both return and absolute return time series. The analysis is conducted over a rolling three-year window, with the window advanced by one trading day at each step to capture the temporal evolution of $h(q)$. For all assets except BTC/USD, the three-year window is defined as 750 trading days (250 days per year). Since BTC/USD is traded continuously, its annual length is set to 365 days. The multifractal strength is quantified by $\Delta h(5)$, representing the difference between $h(-5)$ and $h(5)$.

\begin{figure}[H]
\centering
\includegraphics[width=9.5cm]{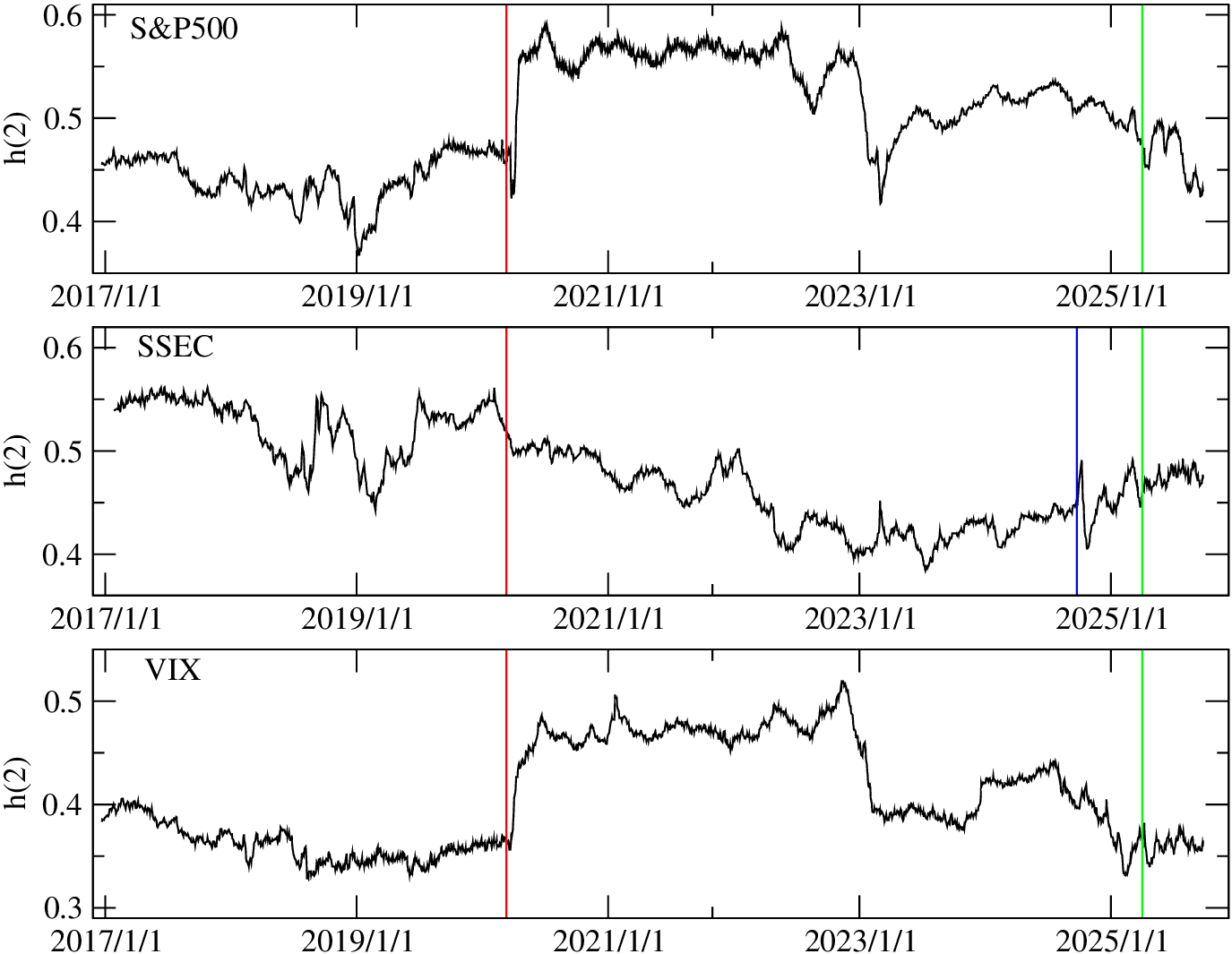}
\caption{Time evolution of the Hurst exponent $h(2)$ for asset returns. Top: S\&P 500; middle: SSEC; bottom: VIX. 
Red line: World Health Organization's declaration of the COVID-19 pandemic (March 11, 2020); 
blue line: Chinese government's announcement of a comprehensive economic stimulus package (September 24, 2024); 
green line: implementation of Trump-era tariffs (April 2, 2025).}
\label{fig:figure1}
\end{figure}

\begin{figure}[h]
\centering
\includegraphics[width=9.5cm]{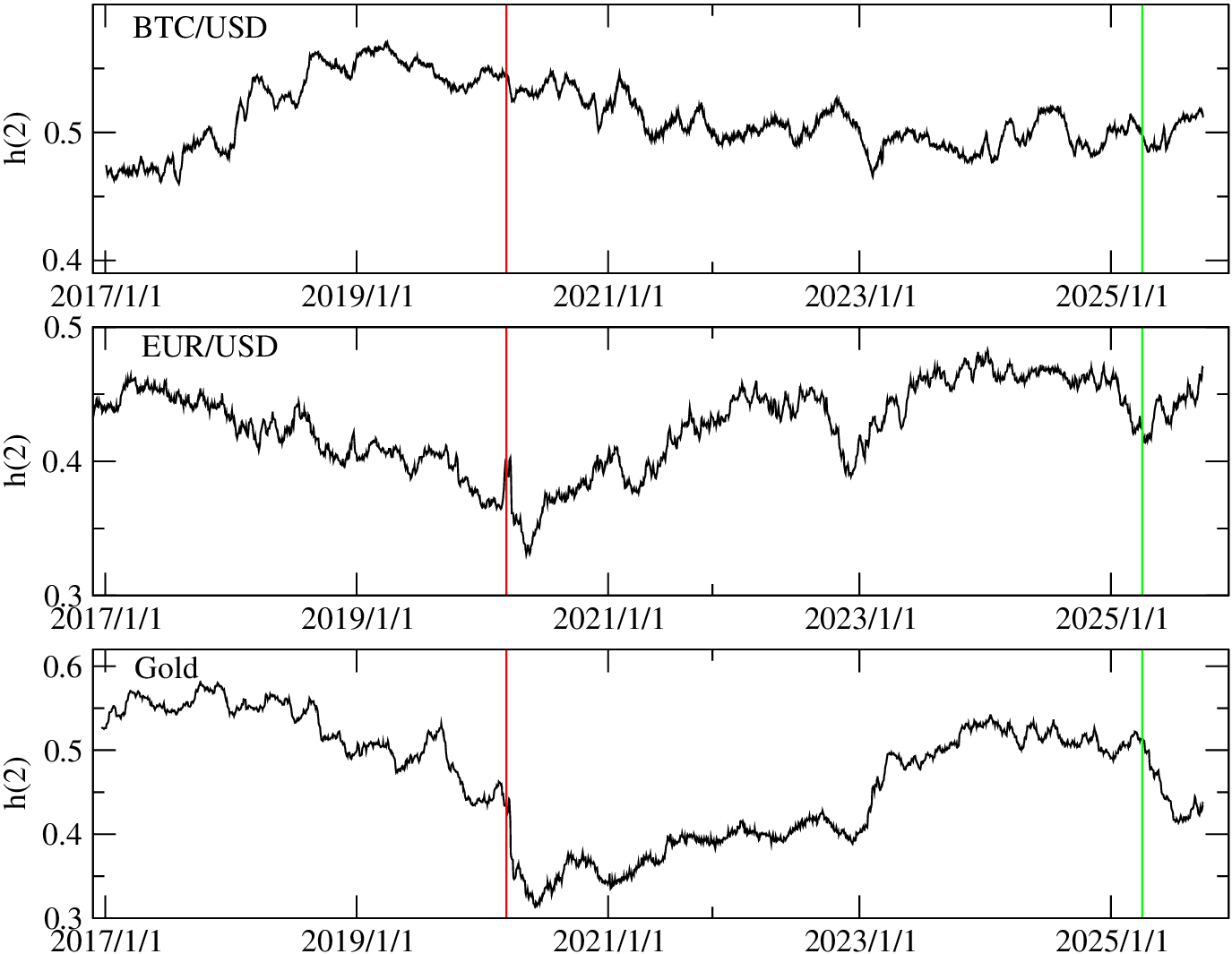}
\caption{Time evolution of the Hurst exponent $h(2)$ for asset returns. Top: BTC/USD; middle: EUR/USD; bottom: Gold. 
The colored lines indicate the same events as in Figure~\ref{fig:figure1}.}
\end{figure}

Figures 4 and 5 present the time evolution of the Hurst exponent $h(2)$ derived from return time series. Notably, the S\&P 500 and VIX exhibit immediate responses to the COVID-19 pandemic declaration, whereas the SSEC and BTC/USD show no discernible reaction. For EUR/USD and Gold, $h(2)$ initially declines following the pandemic announcement but subsequently trends upward, indicating partial recovery. In comparison, the impact of the Trump tariff announcement is relatively subdued. Post-announcement, $h(2)$ decreases for the S\&P 500 and Gold, while BTC/USD and EUR/USD display slight increases.


\begin{figure}[H]
\centering
\includegraphics[width=9.5cm]{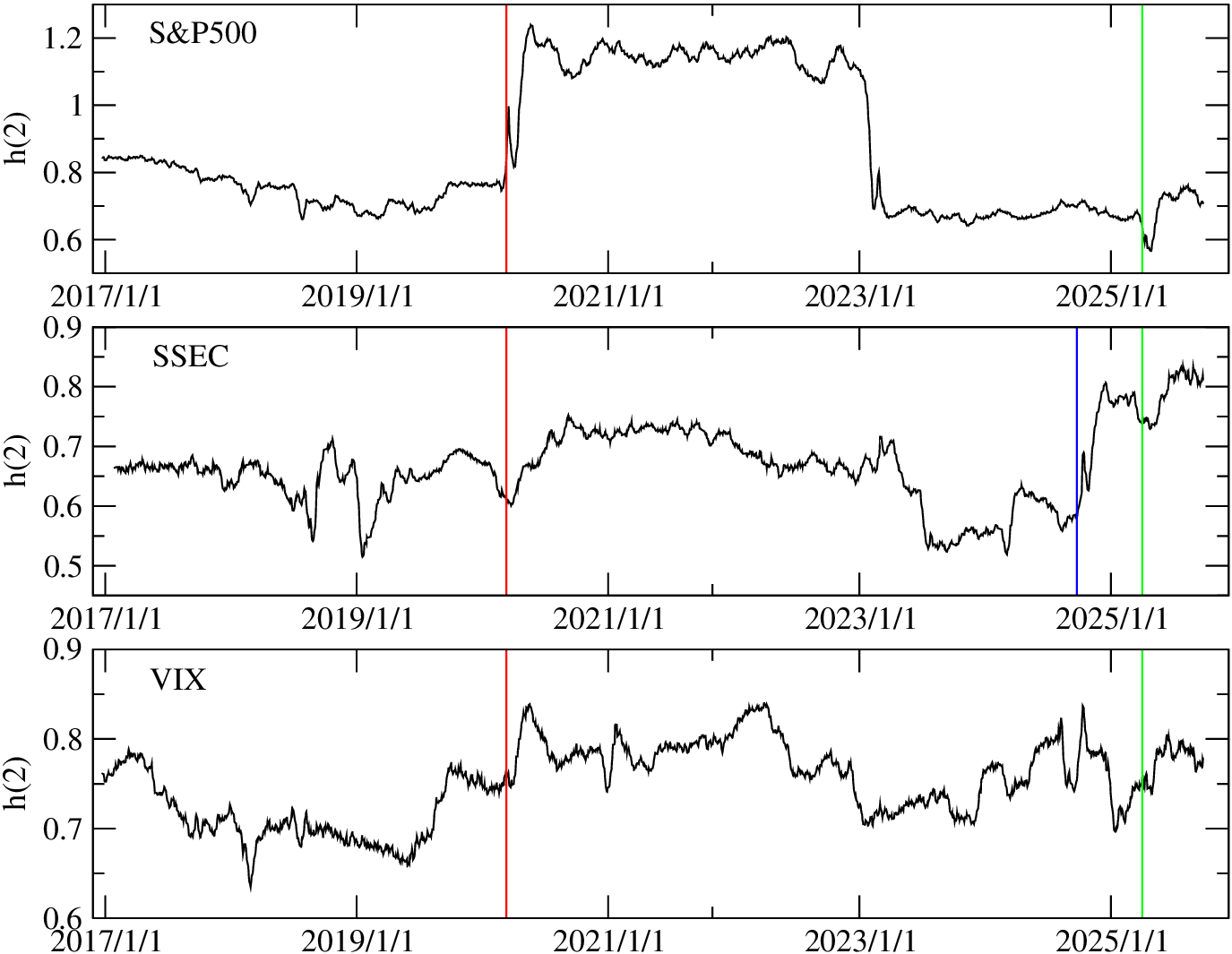}
\caption{Time evolution of the Hurst exponent $h(2)$ for absolute returns. Top: S\&P 500; middle: SSEC; bottom: VIX. 
The colored lines indicate the same events as in Figure~\ref{fig:figure1}.}
\label{fig:figure3}
\end{figure}

\begin{figure}[H]
\centering
\includegraphics[width=9.5cm]{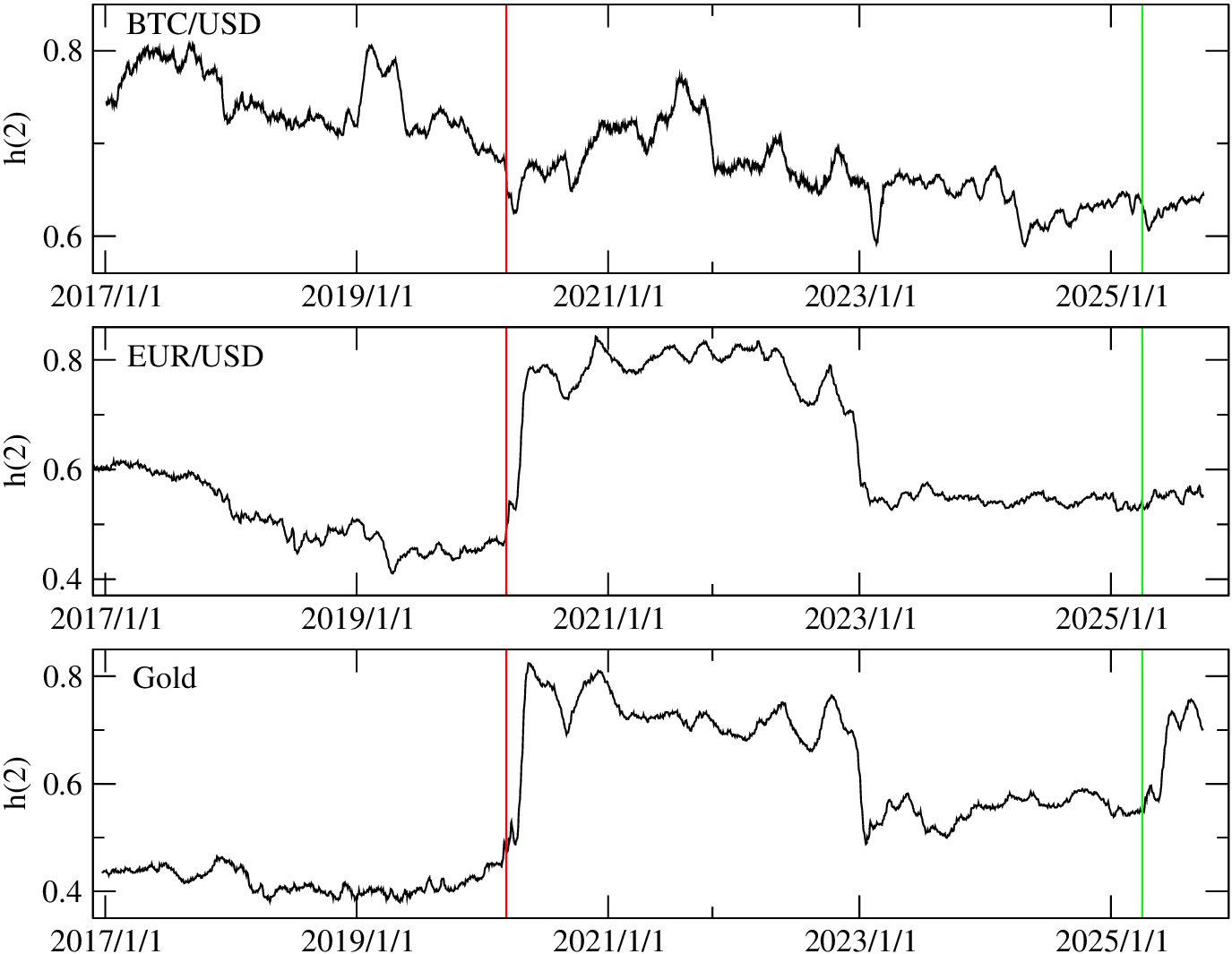}
\caption{Time evolution of the Hurst exponent $h(2)$ for absolute returns. Top: BTC/USD; middle: EUR/USD; bottom: Gold. 
The colored lines indicate the same events as in Figure~\ref{fig:figure1}.}
\label{fig:figure4}
\end{figure}

Figures 6 and 7 depict the temporal variation of $h(2)$ calculated from absolute return series. In contrast to the raw return results, the S\&P 500, EUR/USD, and Gold demonstrate pronounced sensitivity to the pandemic declaration. 
Furthermore, an increase in $h(2)$ of the SSEC has also been observed following the declaration of the pandemic.
The effect of the Trump tariffs remains limited, mirroring the return-based findings; however, with the exception of EUR/USD, $h(2)$ increases following the announcement. For the SSEC, a marked rise in $h(2)$ is observed in response to the Chinese government's economic stimulus measures, as indicated by the blue line.

\begin{figure}[H]
\centering
\includegraphics[width=9.5cm]{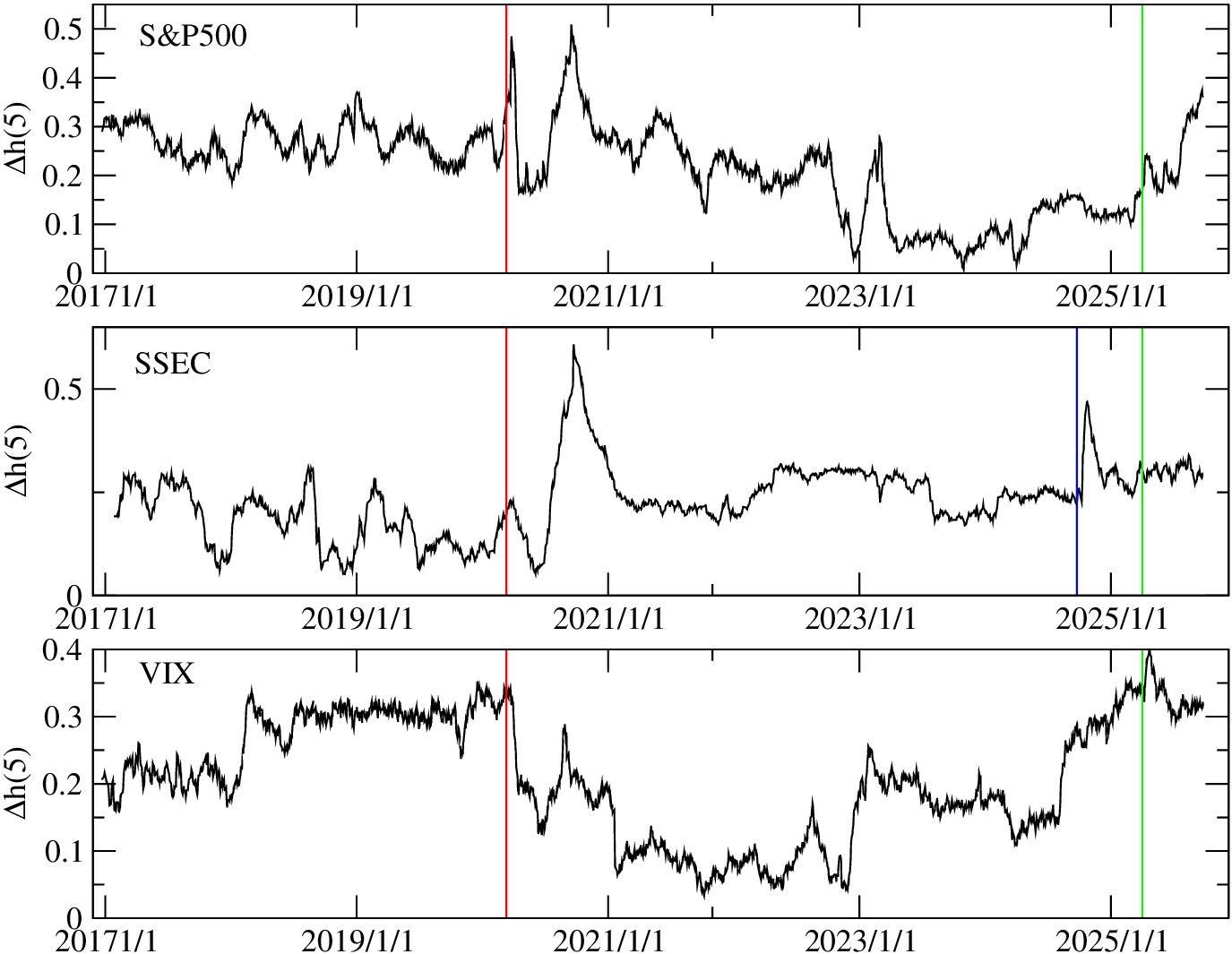}
\caption{Time evolution of the multifractal strength $\Delta h(5)$ for returns. Top: S\&P 500; middle: SSEC; bottom: VIX. 
The colored lines indicate the same events as in Figure~\ref{fig:figure1}.}
\label{fig:figure5}
\end{figure}

\begin{figure}[H]
\centering
\includegraphics[width=9.5cm]{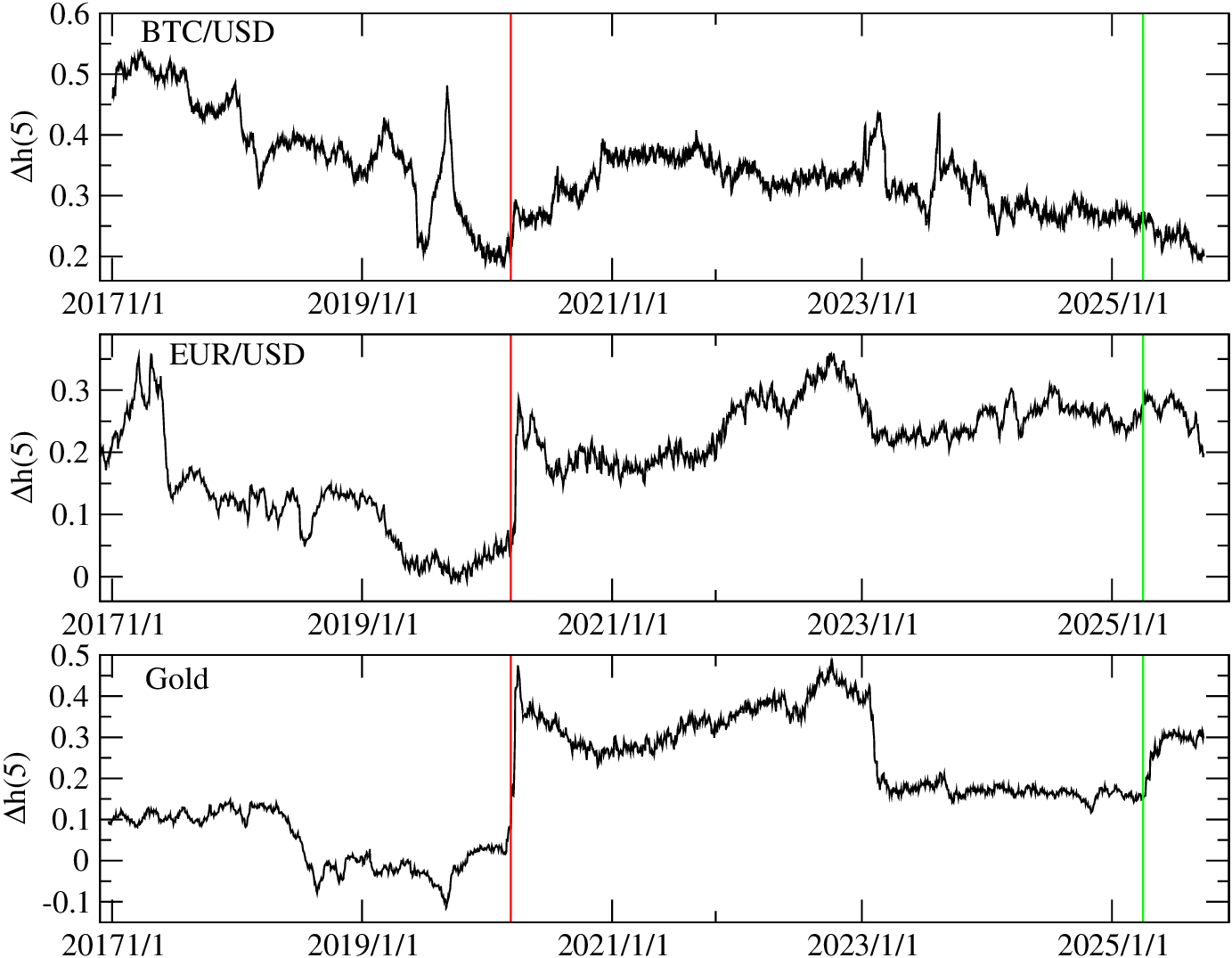}
\caption{Time evolution of the multifractal strength $\Delta h(5)$ for returns. Top: BTC/USD; middle: EUR/USD; bottom: Gold. 
The colored lines indicate the same events as in Figure~\ref{fig:figure1}.}
\label{fig:figure6}
\end{figure}

Figures 8 and 9 illustrate the evolution of multifractal strength $\Delta h(5)$ based on return time series, while Figures 10 and 11 show the corresponding dynamics derived from absolute returns. Although fluctuations in $\Delta h(5)$ are substantial and not always attributable to specific events, several assets exhibit clear responses to the COVID-19 pandemic. For instance, all assets except the SSEC show immediate shifts in return-based multifractal strength following the pandemic declaration. Similarly, in the absolute return-based analysis, significant changes are observed in the S\&P 500, BTC/USD, EUR/USD, and Gold.

\begin{figure}[h]
\centering
\includegraphics[width=9.5cm]{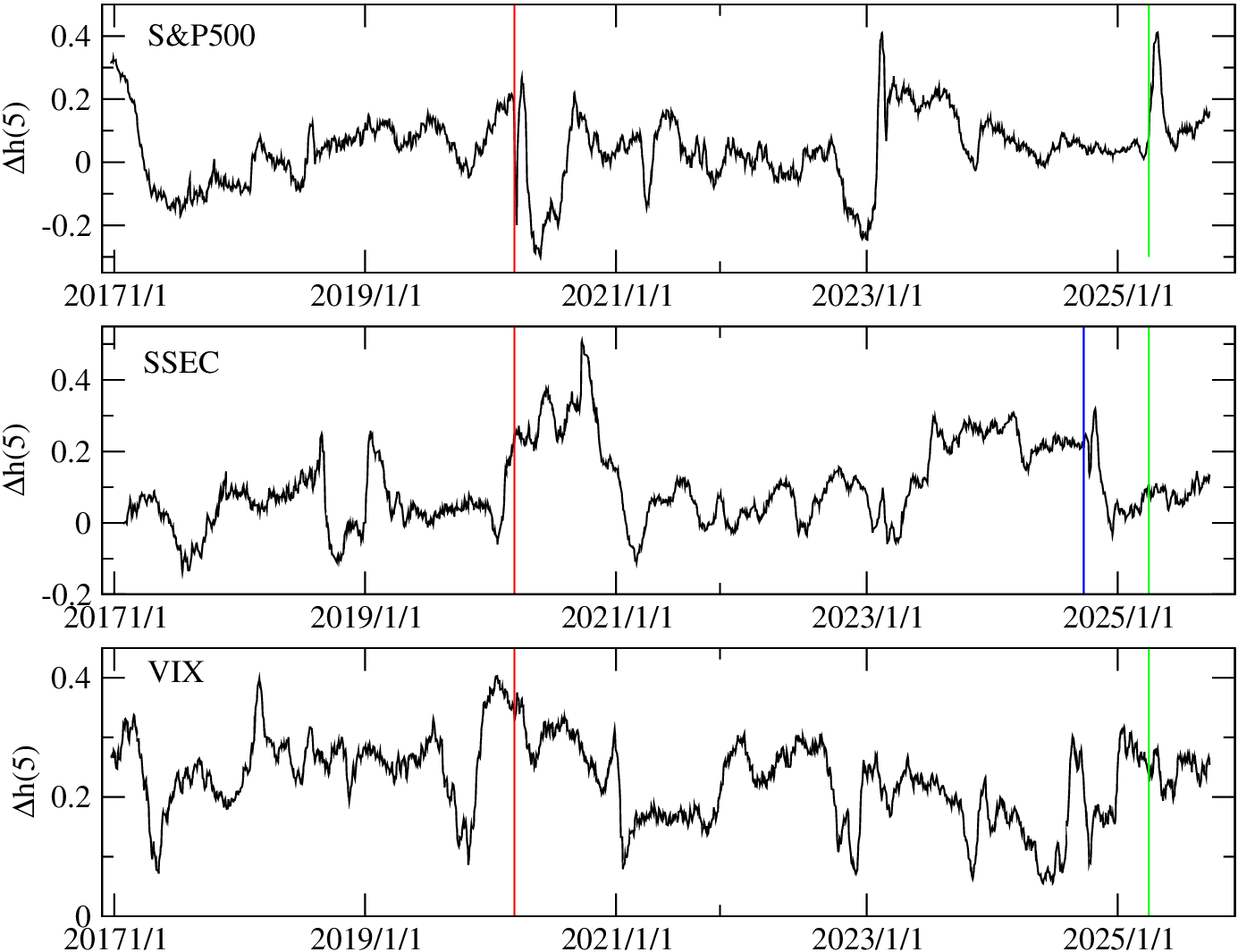}
\caption{Time evolution of the multifractal strength $\Delta h(5)$ for absolute returns. Top: S\&P 500; middle: SSEC; bottom: VIX. 
The colored lines indicate the same events as in Figure~\ref{fig:figure1}.}
\label{fig:figure7}
\end{figure}

\begin{figure}[h]
\centering
\hspace{-3pc}
\includegraphics[width=9.5cm]{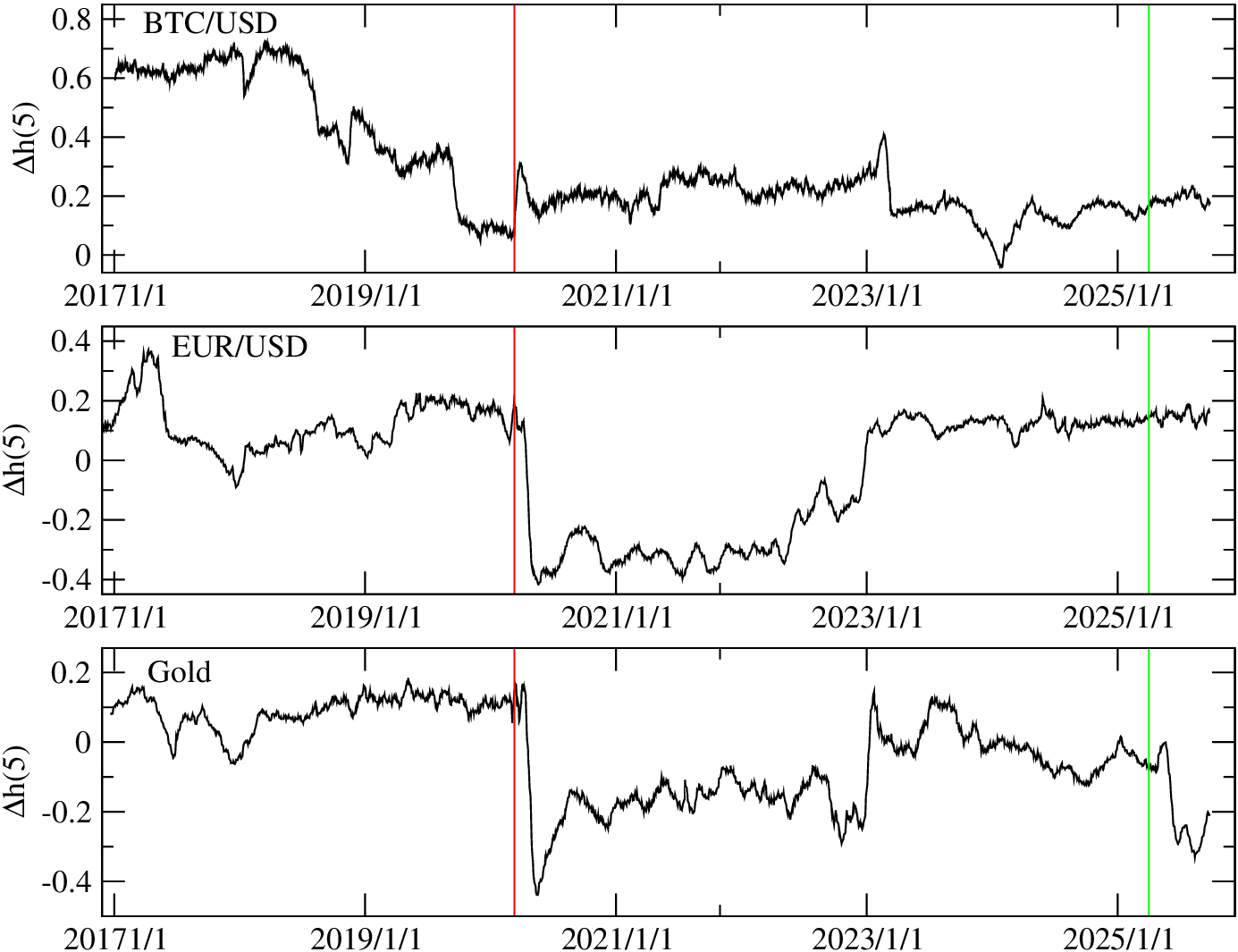}
\caption{Time evolution of the multifractal strength $\Delta h(5)$ for absolute returns. Top: BTC/USD; middle: EUR/USD; bottom: Gold. 
The colored lines indicate the same events as in Figure~\ref{fig:figure1}.}
\label{fig:figure8}
\end{figure}

\begin{figure}[h]
\centering
\includegraphics[width=9.5cm]{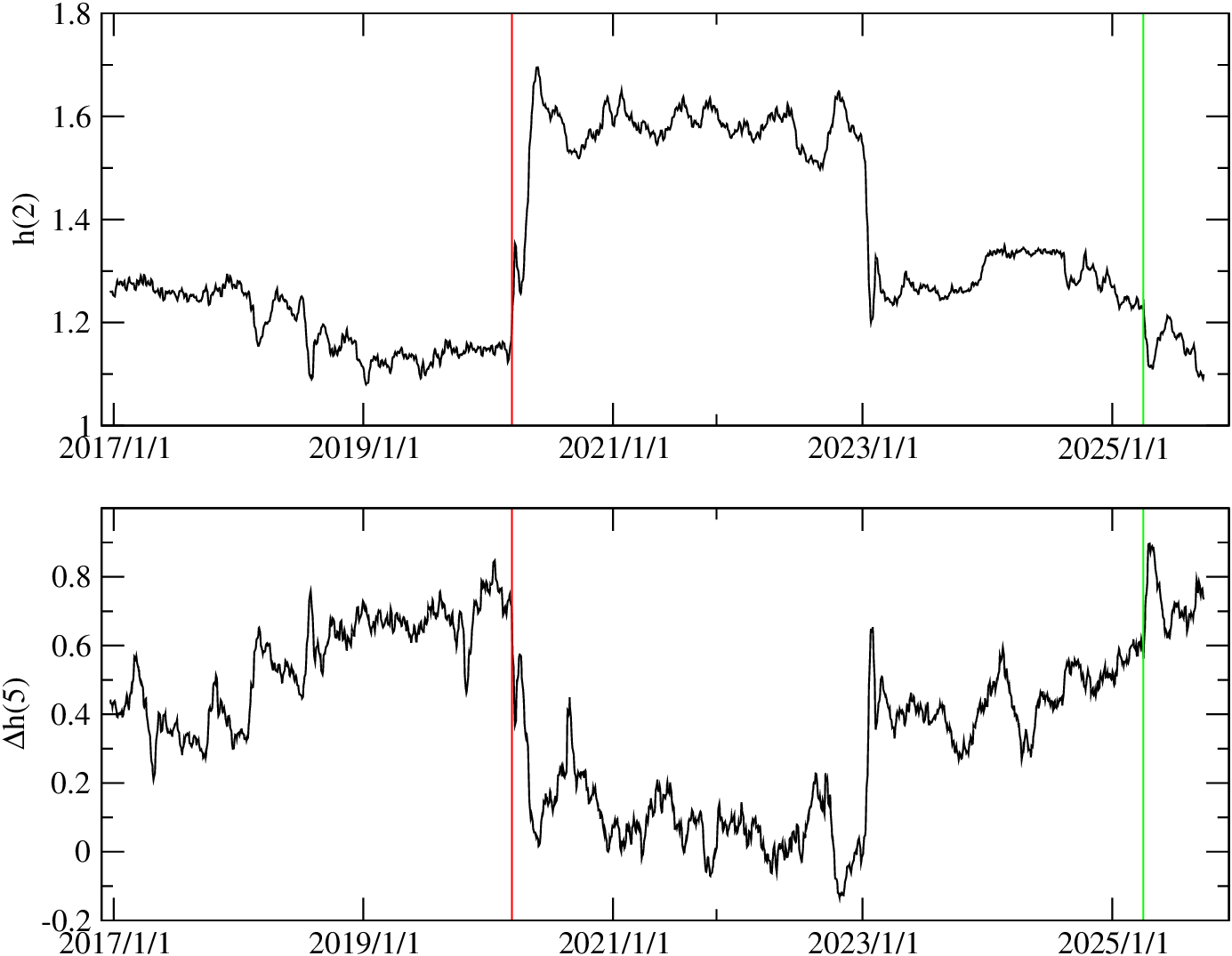}
\caption{Time evolution of the Hurst exponent and multifractal strength for VIX price time series. 
The colored lines indicate the same events as in Figure~\ref{fig:figure1}.}
\label{fig:figure9}
\end{figure}

Consistent with the $h(2)$ results, the influence of the Trump tariffs on multifractal strength is relatively minor. Nonetheless, observable changes are present in the S\&P 500 and Gold time series. The SSEC remains largely unaffected by both global events, although a distinct response to the domestic stimulus policy announced on September 24, 2024, is evident.

The multifractal strength of EUR/USD and Gold for absolute returns shifts to negative values during the pandemic, implying $h(-5) < h(5)$ and suggesting that $h(q)$ becomes an increasing function of $q$ under extreme market conditions.

It is well established that the Hurst exponent $h(2)$ for stock return series typically approximates 0.5, consistent with a random walk. However, as shown in Figure 4, the VIX return series exhibits $h(2) < 0.5$ during non-pandemic periods, indicating anti-persistent behavior. This is characteristic of volatility indices, and aligns with prior empirical findings that volatility-based time series often display rough volatility dynamics \cite{gatheral2018volatility,bennedsen2022decoupling,livieri2018rough,takaishi2020rough,floc2022roughness,brandi2022multiscaling,cont2024rough,takaishi2025multifractality}. The anti-persistence in VIX returns has also been pointed out in \cite{bariviera2023disentangling}.
Given that absolute returns serve as a proxy for volatility, it is reasonable to expect the VIX price series to exhibit similar statistical properties. Figure 12 confirms this, showing that the temporal evolution of $h(2)$ and $\Delta h(5)$ in the VIX price time series closely mirrors that of the absolute return series, particularly in comparison with the S\&P 500.

\section{Conclusion}

This study investigated the impact of the Trump-era tariff policy on financial market efficiency by applying multifractal detrended fluctuation analysis (MFDFA) to return and absolute return time series across six major financial assets. By estimating the generalized Hurst exponent $h(q)$ and multifractal strength $\Delta h(5)$, we captured temporal variations in market dynamics in response to two significant global events: the COVID-19 pandemic and the implementation of the Trump tariffs.

Our findings reveal that the COVID-19 pandemic induced substantial shifts in both the Hurst exponent and multifractal strength, particularly in the S\&P 500, BTC/USD, EUR/USD, and Gold. These changes reflect heightened market instability and structural alterations in efficiency during periods of systemic stress. In contrast, the impact of the Trump-era tariff announcement was comparatively modest, though still discernible across all examined time series.

The Chinese market index (SSEC) remained largely insulated from both global events, with the exception of a pronounced response to domestic economic stimulus measures announced on September 24, 2024. This suggests that local policy interventions may exert a more immediate influence on market structure than external geopolitical shocks.

Additionally, the observed shift to negative multifractal strength in the absolute return series of EUR/USD and Gold during the pandemic implies a reversal in the ordering of $h(-5)$ and $h(5)$, indicating that the general Hurst exponent $h(q)$ became an increasing function of $q$. This phenomenon underscores the sensitivity of multifractal properties to extreme market conditions.

Finally, the anti-persistent behavior observed in the VIX return time series, characterized by $h(2) < 0.5$, is consistent with prior empirical evidence on return volatility and supports the classification of such dynamics within the rough volatility framework. Furthermore, the statistical similarity between the VIX price time series and the absolute return series further substantiates this interpretation.

Overall, our results highlight the utility of multifractal analysis in detecting nuanced structural shifts in market efficiency and provide a comparative lens through which to assess the financial impact of geopolitical and health-related shocks.

\section*{Acknowledgements}
The numerical calculations in this study were carried out using the computer facilities of the Yukawa Institute for Theoretical Physics and the Institute of Statistical Mathematics. 


\bibliographystyle{unsrt}
\bibliography{Trump2-Springerv2.bbl}

\end{document}